\documentclass[12pt,preprint]{aastex}

\begin{document}


\title{GALAXY DOWNSIZING AND THE REDSHIFT EVOLUTION OF OXYGEN AND NITROGEN ABUNDANCES: ORIGIN OF THE SCATTER IN THE N/H--O/H DIAGRAM} 

\author{Leonid S. Pilyugin}
\affil{ Main Astronomical Observatory
                 of National Academy of Sciences of Ukraine,
                 27 Zabolotnogo str., 03680 Kiev, Ukraine}
\email{pilyugin@mao.kiev.ua}
\and
\author{Trinh X. Thuan}
\affil{Astronomy Department, University of Virginia, 
P.O. Box 400325, Charlottesville, VA 22904-4325} 
\email{txt@virginia.edu}

\begin{abstract}
The oxygen and nitrogen abundance evolutions with redshift of   
emission-line galaxies in the Sloan Digital Sky Survey are 
considered for four intervals 
of galaxy stellar masses, ranging from 10$^{11.3}$ $M_\sun$ to 
10$^{10.2}$ $M_\sun$.  
We have measured their line fluxes and 
derived the O and N abundances using recent calibrations. 
The evolution of O and N abundances with redshift clearly shows the 
galaxy downsizing effect, where  
enrichment (and hence star formation) ceases 
in high-mass galaxies at earlier times and  
shifts to lower-mass galaxies at later epochs.   
The origin of the scatter in the N/H -- O/H diagram has been examined. 
The most massive galaxies, where O and N enrichment and star formation 
has already 
stopped, occupy a narrow band in the  N/H -- O/H diagram, defining 
an upper envelope. 
The less massive galaxies which are still undergoing star formation 
at the current epoch are shifted downwards, 
towards lower N/H values in the N/H -- O/H diagram.  
This downward shift is caused by the time delay between N and O 
enrichment. This time delay together with the different star formation 
histories in galaxies is responsible for the large scatter in the N/H -- O/H 
diagram. 
\end{abstract}


\section{INTRODUCTION}

Oxygen and nitrogen are key elements in the study of the 
chemical evolution of 
galaxies. The N/O -- O/H (or N/H -- O/H) diagram has
 been considered by many authors
\citep[e.g.][]{edmundspagel1978,matteucci1985,pilyugin1992,pilyugin1993,henryetal2000,gavilanetal2006,izotov2006}. 
A prominent feature of this diagram is that, in  
its high-metallicity part (12+log(O/H) $\ga$ 8.3),   
the N/H abundance ratio shows a large scatter at a 
fixed value of the O/H abundance ratio 
\citep{henryetal2000,pilyuginetal2003,lopezsanchez2010}. 
The N/H value for a given O/H contains important information 
about the heavy element enrichment history of 
a galaxy, and consequently, about its star formation history. 
Therefore, an explanation of the origin of the scatter in the N/H -- O/H 
diagram is key to understanding the evolution of galaxies. 

Two main explanations are presently considered to account for the scatter
in normal spiral galaxies: 
1) the time delay between nitrogen and oxygen enrichment;  
and 2) the local enrichment in nitrogen by Wolf-Rayet (WR) stars.
Concerning the first explanation, \citet{edmundspagel1978} 
have noted that, 
due the fact that O and N are produced in stars of different masses,  
there can be a significant time delay between the release of O, produced in high-mass stars,  and 
that of N, produced in intermediate-mass stars,
 into the interstellar medium (ISM). The N/O ratio of a galaxy then becomes an 
indicator of the time that has elapsed since the 
last episode of star formation. 
Current stellar evolution models predict that N is 
mainly manufactured and ejected 
into the ISM by intermediate-mass stars with masses 
greater than 3-4$M_\sun$ \citep{renzinivoli1981,vandenhoek1997,marigo2001,gavilanetal2005}, 
although massive stars can contribute, at least at low metallicities 
\citep{chiappini2005}.  
This suggests that the time delay between N and O enrichment in galaxies
is  250 -- 400 Myr, depending on the adopted stellar mass--lifetime 
relation \citep{romanoetal2005}. 
Properties of the observed N/O -- O/H diagram can generally be   
reproduced by 
chemical evolution models 
of galaxies by taking into account  a time delay  
\citep{matteucci1985,pilyugin1992,pilyugin1993,pilyugin1999,henryetal2000,contini2002,gavilanetal2006}.  
The exact value of the time delay is however still a subject of debate  
\citep{pilyuginetal2003,richer2008,thuanetal2010}. 

Concerning the second explanation, 
it is known \citep[][and references therein]{lopezsanchez2010} 
that some galaxies  
with WR features have a high N/O ratio. 
This suggests that the ejecta of WR 
stars may locally enrich the ISM in N. 
\citet{henryetal2000} noted that most points in the N/O -- O/H diagram 
cluster at relatively low N/O values. This led them 
to conclude that the lower envelope is the ``equilibrium'' 
unperturbed locus and that the observed 
scatter is the result of intermittent increases in N, 
caused by the local contamination by WR stars or luminous blue variables. 
\citet{izotov2006} also arrived at the same conclusion. 

Selective heavy element loss through enriched galactic winds may also 
introduce scatter in the N/O ratios. It is believed, however, that 
galactic winds play 
an important role in the chemical evolution of only 
dwarf galaxies,  not of giant spirals \citep[e.g.][]{pilyuginetal2004}.

In recent years, the number of good-quality 
spectra of emission-line galaxies has 
increased dramatically due to the completion of several large spectral surveys, 
and in particular of the Sloan Digital Sky Survey (SDSS) \citep{yorketal2000}.
This opens the 
possibility of using the SDSS spectral base to study the evolution 
of O and N abundances in galaxies in the redshift 
range z $\la$ 0.4 \citep[e.g.][]{tremontietal2004}.  In a  
previous study \citep{thuanetal2010}, we have 
considered the evolution 
of O and N abundances in galaxies with different stellar masses. 
We have found clear evidence for galaxy downsizing, where the 
sites of active star formation and hence of metal enrichment shift from 
high-mass galaxies at early cosmic times to lower-mass systems at later epochs 
\citep{cowie1996}. All enrichment ceases in the most massive galaxies at 
late cosmic times. 
This downsizing effect provides a remarkable opportunity 
for clarifying the origin of the scatter in the N/H -- O/H diagram. 
Indeed, if the scatter is caused 
by a time delay  between O and N enrichment in galaxies,
 then it should be minimized 
for the most massive galaxies where 
enrichment has stopped. On the other hand, one should expect that the  
N/H ratios of the less massive galaxies 
(those with significant star formation at the current epoch) to 
be shifted towards lower values relative to the massive 
galaxies
in the N/H -- O/H diagram, because there is a time delay between N and O 
enrichment and N has not yet been released in the ISM.

We examine in this paper whether the time delay is in fact responsible for the 
large scatter in the N/H -- O/H diagram.   
Our previous study 
was based on the MPA/JHU catalogs of automatic line flux measurements of 
the SDSS spectra (see  
 \citet{brinchmannetal2004,tremontietal2004} and other publications of those authors).
The accuracy of such 
automatic line flux measurements seems is not good enough 
for investigating the origin of scatter in the N/H -- O/H diagram. 
We have thus decided to manually measure the line fluxes of the SDSS spectra. 
This provides us with more accurate line flux measurements, 
especially for spectra of objects with
the largest redshifts, those with redshifts between 0.3 and 0.4.

\section{SAMPLE SELECTION}
 
We first extract from the MPA/JHU catalogs 
four subsamples of emission-line galaxies, each covering a different  
interval of galaxy stellar mass, centered respectively on the mass values 
$M_S$ = 10$^{11.3}$$M_\sun$,  10$^{11.0}$$M_\sun$,   
10$^{10.6}$$M_\sun$,  and 10$^{10.2}$$M_\sun$. 
To have a reasonable number of galaxies in each subsample and in 
order for the galaxies to cover the whole redshift range in 
a roughly uniform fashion, 
the mass interval $dM$ was chosen to be small and to vary from 0.0015 
to 0.10 dex, depending on the redshift and $M_S$.
For each mass interval, 
we have extracted from the MPA/JHU catalogs emission-line 
galaxies with automatic measurements of fluxes in the H$\beta$, H$\alpha$, 
[O\,{\sc ii}]$\lambda \lambda$3727,3729, 
[O\,{\sc iii}]$\lambda$5007, [N\,{\sc ii}]$\lambda$6584 emission lines. 
We impose the additional restrictions that the galaxies 
have equivalent widths 
$EW$(H$\beta$) $>$ 10 {\AA} (this criterion applies only to galaxies 
with redshifts $<$ 0.3, as we did not wish to reduce 
further the very small number of galaxies with $z$ $>$ 0.3),
$EW$([O\,{\sc ii}]$\lambda \lambda$3727,3729) $>$ 3 {\AA}, 
$EW$([O\,{\sc iii}]$\lambda$5007) $>$ 3 {\AA}, and 
$EW$([N\,{\sc ii}]$\lambda$6584) $>$ 2 {\AA}. These additional restrictions 
insure that the chosen galaxies 
have a reasonably high star formation rate, giving rise to reasonably 
strong emission lines that can be measured with good accuracy.  
 
Each so chosen spectrum was then examined visually and  
the noisy spectra were rejected. 
The line fluxes of the remaining galaxies were then measured 
with IRAF  \footnote{IRAF is distributed by National Optical Astronomical Observatories,
which are operated by the Association of Universities for Research in Astronomy, Inc., 
under cooperative agreement with the National Science Foundation.}.
The [O\,{\sc iii}]$\lambda$5007)/H$\beta$ vs [N\,{\sc ii}]$\lambda$6584)/H$\alpha$ 
diagram was then used 
to reject AGNs \citep{baldwin1981}, with 
the dividing line between H\,{\sc ii} regions 
ionized by star clusters and AGNs taken from \cite{kauffmannetal2003}.

The wavelength range of the SDSS spectra is 3800 {\AA} -- 9300 {\AA}, so that
for nearby galaxies with redshift z $\la$ 0.023, the 
[O\,{\sc ii}]$\lambda$3727+$\lambda$3729 emission line is out of the 
observed range. Thus, all galaxies in our total sample have redshifts greater than   
$\sim$ 0.023.
For distant galaxies with redshift z $\ga$ 0.33,  
the [S\,{\sc ii}]$\lambda$$\lambda$ 6717,6731 emission lines are out of the observed range. 
The sulfur line intensities are usually used as indicators of the 
electron density.  
They serve also to distinguish between hot and warm 
H\,{\sc ii} regions, 
in the framework of the ON calibrations of \citet{pilyuginetal2010}. 
Since the sulfur lines are not present, we will assume the 
electron density to be equal to 100 cm$^{-3}$ 
and that our sample does not contain hot H\,{\sc ii} regions 
(i.e. those with 12+log(O/H) $\la$ 8.0). 
The redshift $z$ and stellar mass $M_S$ of each galaxy were taken from 
the MPA/JHU catalogs.
When the stellar mass of a galaxy with redshift $z$ $>$ 0.3 is not available 
in the MPA/JHU catalogs, we estimated it from its 
absolute magnitude $M_z$, its colour $m_z$--$m_r$ and its H$\beta$ 
equivalent width, using the relation 
\begin{eqnarray}
       \begin{array}{lll}
\log M_S  & =   &  -0.459\,M_z - 0.775\,(m_z-m_r) + 0.363 +0.083 \log EW({\rm H}\beta)     \\ 
          & for &  (m_z - m_r) < - 0.3                                                   \\
          &     &                                                    \\
\log M_S  & =   &  -0.61\,M_z + 0.43\,(m_z-m_r) -2.58 - 0.24 \log EW({\rm H}\beta)     \\ 
          & for &  (m_z - m_r) > - 0.3                                                   \\
     \end{array}
\label{equation:ms}   
\end{eqnarray}
 where the galaxy stellar mass $M_S$ is in units of solar masses. 
This relation was derived by fitting the data  
for galaxies in the redshift interval 0.3 $<$  $z$ $<$ 0.4 with available  
stellar masses in the MPA/JHU catalogs. 
We caution that this relation is not 
a general relation for the determination of galaxy stellar masses.
It applies only within the small particular redshift interval defined above.
We stress also that the derived masses serve only to select 
galaxies with various masses to populate our different subsamples and are  
not used in any calculation.

Our final database consists of  
221 spectra of galaxies with $M_S$ = 10$^{11.3}$$M_\sun$ (with 74 objects with $M_S$ derived from Eq. 1), 
259 spectra of galaxies with $M_S$ = 10$^{11.0}$$M_\sun$ (60 objects with $M_S$ from Eq. 1),  
244 spectra of galaxies with $M_S$ = 10$^{10.6}$$M_\sun$ (32 objects with $M_S$ from Eq. 1), and  
152 spectra of galaxies with $M_S$ = 10$^{10.2}$$M_\sun$ (7 objects with $M_S$ from Eq. 1). 
The measured emission-line fluxes are then corrected for interstellar 
reddening using the theoretical H$\alpha$ to H$\beta$ ratio,
 in the same way as in \citet{thuanetal2010}. 
O and N abundances are then estimated for each galaxy,
using the recent ON calibrations of \citet{pilyuginetal2010}.
These calibrations give both O and N abundances over the whole 
metallicity range with a satisfactory precision. 
\citet{pilyuginetal2010} found that 
the mean differences between O abundances determined from these calibrations and from the 
direct $T_e$ method based on the [O\,{\sc iii}]$\lambda$4363 line 
is only $\sim$0.075 dex for O abundances, 
and only $\sim$0.05 dex for N abundances.

\section{GALAXY DOWNSIZING AND THE ORIGIN 
OF THE SCATTER IN THE N/H -- O/H DIAGRAM}

\begin{figure}
\begin{center}
\resizebox{0.50\hsize}{!}{\includegraphics[angle=0]{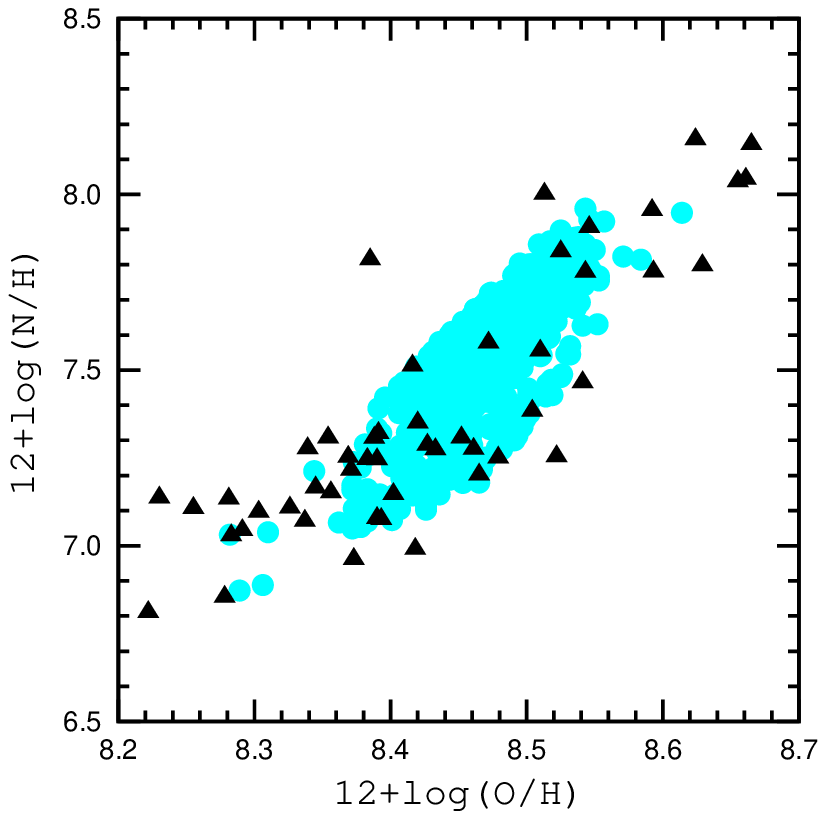}}
\caption{ 
The N/H -- O/H diagram. 
The gray (light-blue in the electronic version) points represent 
individual galaxies 
from our four SDSS subsamples with oxygen and nitrogen abundances derived 
from the ON calibrations of \cite{pilyuginetal2010}.
The dark (black in the electronic version) triangles show H\,{\sc ii} regions 
in nearby galaxies (from a compilation of \cite{pilyuginetal2010} 
with oxygen and nitrogen abundances derived with the $T_e$ method. 
(A color version of this figure is available in the online journal)
}
\label{figure:ohnhte}
\end{center}
\end{figure}

We first check the accuracy of our abundance determinations. 
Fig.\ref{figure:ohnhte} shows the N/H -- O/H diagram. 
The gray points represent galaxies 
in our four SDSS subsamples with O and N abundances derived 
from the ON calibrations.
The dark triangles show H\,{\sc ii} regions 
in nearby galaxies \citep{pilyuginetal2010}, with O and N abundances derived 
with the $T_e$ method, 
commonly thought to be the most accurate one. 
It is seen that the galaxies 
in our SDSS subsamples occupy the same region 
in the N/H -- O/H diagram as 
the H\,{\sc ii} regions in nearby galaxies with O and N abundances 
derived through the $T_e$ method. This confirms that 
our ON-calibration-based O and N abundances are quite reliable. 
The remarkable feature to note in 
the N/H -- O/H diagram is the large scatter in N/H values at a given 
O/H value. This holds as much for SDSS objects as 
for H\,{\sc ii} regions in nearby galaxies. 

\begin{figure}
\begin{center}
\resizebox{0.50\hsize}{!}{\includegraphics[angle=0]{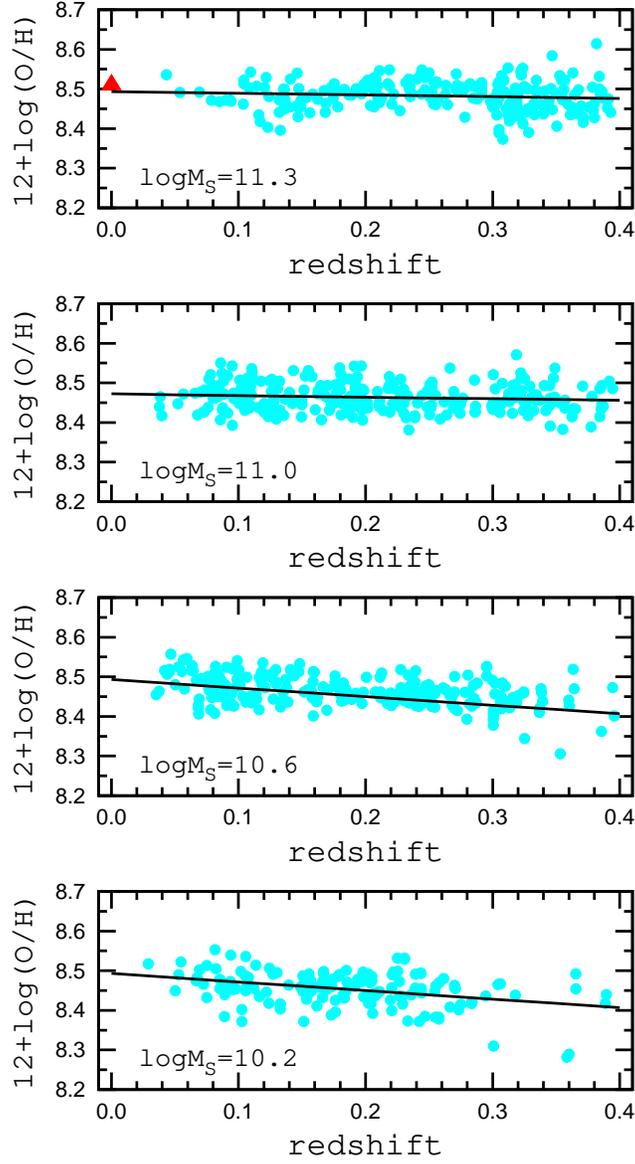}}
\caption{ 
Oxygen abundances as a function of redshift for the four 
subsamples of galaxies with different stellar masses. 
 In each panel, labeled by the galaxy stellar mass, 
the gray (light-blue in the electronic version) points 
represent individual galaxies. 
The solid (black in the electronic version) line is the 
least-squares best fit to those data. 
The dark (red in the electronic version) filled triangle in the top panel 
shows the Orion Nebula.
(A color version of this figure is available in the online journal)
}
\label{figure:zoh}
\end{center}
\end{figure}

We next investigate the redshift evolution of O abundances in each 
of the four galaxy mass ranges (Fig.\ref{figure:zoh}).
In each panel labeled by the galaxies' masses, 
the filled gray  
circles show individual galaxies.  
The solid line is the best least-squares fit to those data. 
Inspection of the upper panel shows that 
there is no systematic variation with redshift of the O abundance in galaxies with 
masses $\sim$ 10$^{11.3}$M$_\sun$  up to $z$ = 0.4. 
 This implies that these galaxies have reached a high 
astration level some 4 Gyr ago, and have been somewhat "lazy" in their 
evolution afterwards. 
The average value of the O abundances for galaxies with stellar 
mass  M$_S$ = 10$^{11.3}$M$_\sun$ is remarkably close to the O abundance 
in the Orion nebula (12+log(O/H) = 8.51) obtained by \citet{estebanetal2004}. It is 
also in good agreement with the O abundances in nearby luminous galaxies 
obtained by \citet{moustakasetal2010}, using the calibration of \cite{pilyuginthuan2005}. 

Comparison between the different panels of Fig.\ref{figure:zoh} shows 
clearly the effect of galaxy downsizing.
Massive galaxies do not show O enrichment: they already reach a 
relatively high O abundance at $z$ = 0.4, 
and that abundance remains nearly constant until $z$ = 0. 
The less massive galaxies do show O enrichment: 
they have lower O abundances 
at $z$ = 0.4, but these increase from $z$ = 0.4 
to $z$ = 0, and at the present epoch, they are nearly the same as 
in massive galaxies.

\begin{figure}
\begin{center}
\resizebox{1.00\hsize}{!}{\includegraphics[angle=0]{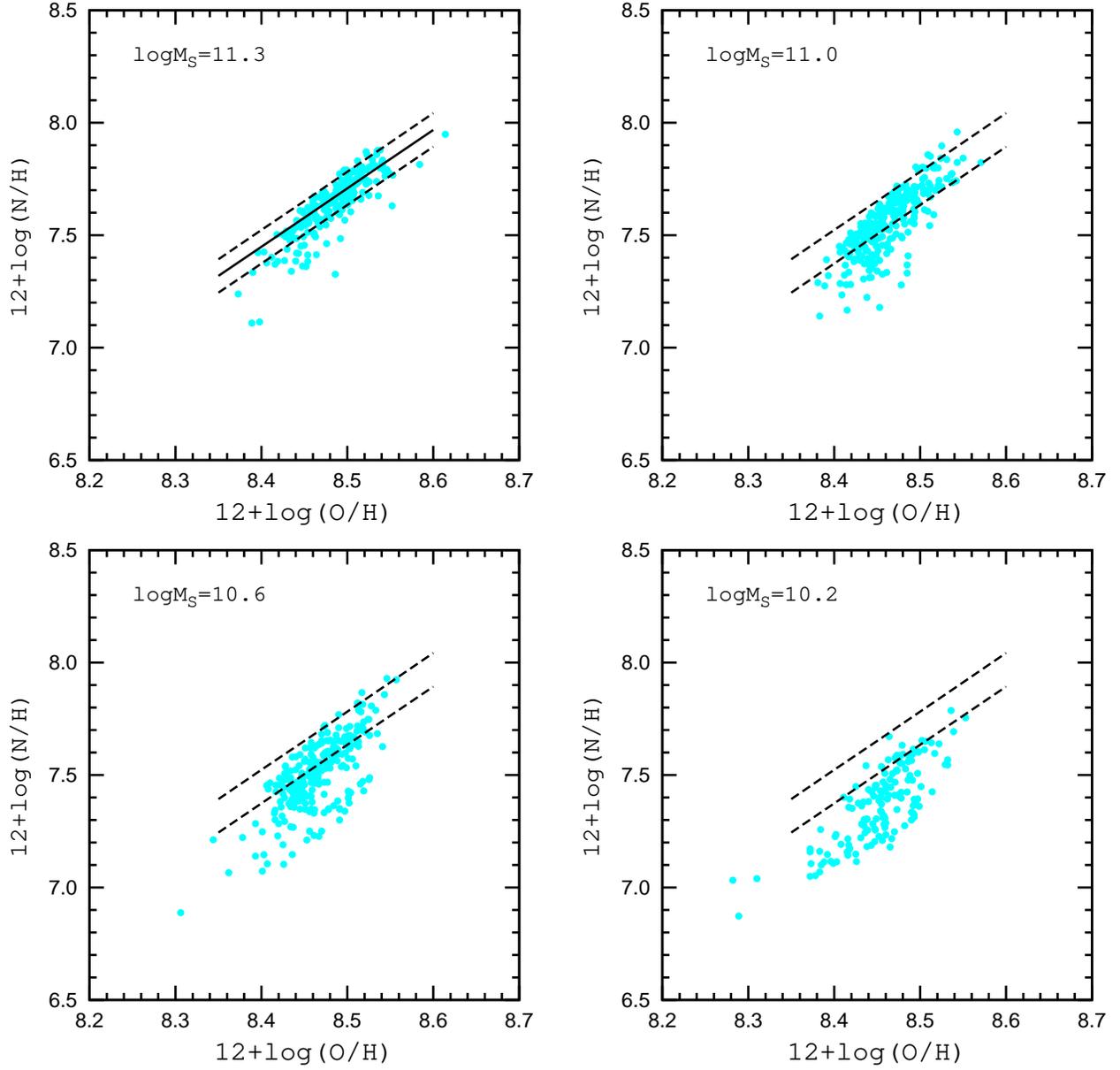}}
\caption{ 
N/H -- O/H diagrams for the four subsamples of galaxies with different 
stellar masses.
In each panel, labeled by the galaxy stellar mass,  
the gray (light-blue in the electronic version) points represent 
individual galaxies. 
The solid (black in the electronic version) line in the left upper panel 
is the derived  N/H -- O/H relation (equation 2) for the 
most massive galaxy subsample, with masses $\sim$ 10$^{11.3}$M$_\sun$.
The dashed (black in the electronic version) lines show 
the $\pm$2$\sigma$ shifts from this relation. 
These dashed lines are reproduced in all other panels.
 (A color version of this figure is available in the online journal)
}
\label{figure:ohnh}
\end{center}
\end{figure}

We next investigate the origin of the scatter in the N/H -- O/H diagram. 
The upper left panel of Fig.\ref{figure:ohnh} shows that diagram 
for the subsample 
that contains our most massive galaxies, those
 with masses $\sim$ 10$^{11.3}$M$_\sun$.
 It is seen that the scatter in the N/H -- O/H  diagram 
for this subsample of galaxies is small, especially for galaxies with $z$ $\la$ 0.3 (see Fig. 4, to be discussed later). 
Using this subsample, 
we have derived the following relation between N/H and O/H abundances 
\begin{equation} 
\log ({\rm N/H})   =  2.596 \,(\pm 0.118) \, \log({\rm O/H}) - 14.359 \,(\pm 1.001) . 
\label{equation:nh-oh}    
\end{equation} 
This relation has been derived in an iterative manner. First,   
a least-squares fit was obtained for all data points with $z$ $<$ 0.3. 
Then, objects with deviations larger than 2$\sigma$ were rejected and a new  
least-squares fit derived. The final fit is obtained when, 
after several iterations, the fits for two consecutive ones 
coincide. The scatter in the N/H values relative to the final 
fit is $\sigma$ = 0.037 dex. 
The derived N/H -- O/H relation is shown in the upper left panel of Fig.\ref{figure:ohnh} by 
a solid line, while 
the dashed lines show 
$\pm$2$\sigma$ shifts from that relation. 
It should be noted that this N/H -- O/H relation is derived for a relatively 
small range of O/H  and, consequently, its slope 
cannot be determined with a high precision. 
Therefore, it should not be used for O/H values  
beyond the above specified range. To establish a more general 
 N/H -- O/H relation, a sample of objects covering a more extended 
range in O/H would be needed.    

The above relation links N/H and O/H in galaxies where O and N enrichment 
has already ceased. 
One would expect that galaxies with significant star formation 
in the near past would be displaced 
towards lower N/H values in the N/H -- O/H diagram
relative to this relation,
 because of the time delay between N and O enrichment.
Comparison between the four 
panels of Fig.\ref{figure:ohnh} clearly shows this effect. 
The most massive galaxies define an upper envelope in the N/H -- O/H diagram 
and they do not show significant star formation in the last 4 Gyr 
(Fig.\ref{figure:zoh}).
The lower-mass galaxies are shifted towards lower N/H values 
and they show significant star formation during the same period   
 (Fig.\ref{figure:zoh}). 
This is solid evidence that the differences in N/H for galaxies 
with a given O/H are caused by the time delay between N and O enrichment.
 
\begin{figure}
\begin{center}
\resizebox{0.44\hsize}{!}{\includegraphics[angle=0]{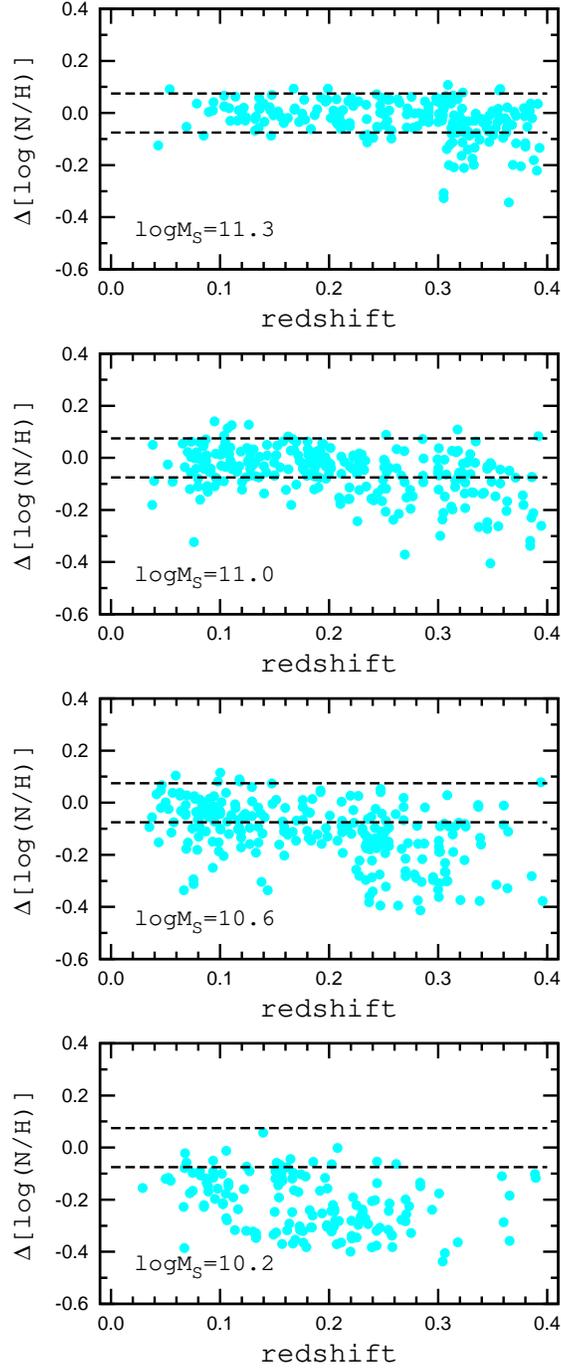}}
\caption{ 
Deviations of the nitrogen abundance from the N/H--O/H relation, 
derived for the most massive galaxy subsample 
(equation 2), as a function of redshift for all four subsamples of galaxies  
with different stellar masses. 
In each panel, labeled by the galaxy stellar mass, 
the gray (light-blue in the electronic version) points represent 
individual galaxies. 
The dashed (black in the electronic version) lines show 
the $\pm$2$\sigma$ deviations from the N/H--O/H relation. 
(A color version of this figure is available in the online journal)
}
\label{figure:zdn}
\end{center}
\end{figure}

Fig.\ref{figure:zdn} shows how  
the deviations of the N abundance from the N/H -- O/H
relation (Equation 2) vary with mass and redshift. 
The gray points 
represent individual galaxies. 
The dashed  
lines show $\pm$2$\sigma$ deviations from that relation. 
The N evolution with redshift in galaxies of different masses 
is a clearly due to a galaxy downsizing effect. As 
the universe ages, the sites of star formation shift from high-mass 
galaxies (log M = 11.3), where star formation and 
O and N enrichment have ceased more than 
4 Gyr ago, to lower-mass galaxies (log M $<$ 11.3) 
where active star formation and O and N enrichment 
are still ongoing. 
Comparison of Figs.\ref{figure:zdn} and \ref{figure:zoh} 
shows that the downsizing effect is more prominent 
for N than for O. This is because N is a 
secondary element while O is a primary one.

To have comparable numbers of galaxies with low redshifts 
(0.02 $<$ $z$ $<$ 0.15) in the highest-mass subsample of galaxies 
($M_S$ = 10$^{11.3}$$M_\sun$) and in the lowest-mass 
subsample ($M_S$ = 10$^{10.2}$$M_\sun$),  we had to 
use a mass interval $dM$ = 0.10 in the first case, and 
a value about 66 times lower,   
$dM$ = 0.0015, in the second case. 
This implies that, at the current epoch,
star formation events occur nearly two orders of magnitude 
more often in low-mass galaxies than in high-mass ones. 
Because of the time delay between O and N enrichment, 
there is a downward shift towards 
lower N/H values for low-mass 
galaxies in the  N/H -- O/H  diagram (Fig.\ref{figure:zdn}). 
This naturally explains the finding of \citet{henryetal2000} that most 
of the local dwarf star-forming galaxies 
 cluster at relatively low N/O values in the N/O -- O/H  diagram.

\section{CONCLUSIONS}

The O and N abundance evolutions with redshift in  
emission-line galaxies 
from the Sloan Digital Sky Survey (SDSS) are investigated. 
To investigate the galaxy downsizing effect, 
we have measured line fluxes of selected galaxies in 
four intervals of galaxy stellar masses, ranging from 10$^{11.3}$ $M_\sun$ to 
10$^{10.2}$ $M_\sun$.
The O and N abundances have been derived using the recent accurate  
calibrations of \citet{pilyuginetal2010}.

We have found that the O and N abundance evolutions with redshift of the 
galaxies in our sample clearly show the galaxy downsizing effect 
\citep{cowie1996}, where  
enrichment (and hence star formation) 
is shifted from high-mass galaxies at earlier cosmic times 
to lower-mass galaxies at later epochs.

We have examined the origin of the scatter in the N/H -- O/H diagram. 
We have found that the most massive galaxies 
where O and N enrichment and star formation has already 
stopped, occupy a narrow band in the N/H -- O/H diagram, defining 
an upper envelope. 
On the other hand, the less massive galaxies with significant star formation 
at the current epoch are shifted downwards, 
towards lower N/H values. 
The scatter of N/H for 
a given O/H is caused by the time delay between N and O 
enrichment and the different star formation 
histories in different galaxies.  

\subsection*{Acknowledgments}

L.S.P. thanks the hospitality of the Astronomy Department of the 
University of Virginia.
He acknowledges the support of the Cosmomicrophysics project of 
the National Academy of Sciences of Ukraine. 
T.X.T. thanks the support of NASA.


\newpage

\newpage

\newpage

\newpage

\begin{thebibliography}{}

\bibitem [Baldwin et al.(1981)]{baldwin1981}
          Baldwin, J.A., Phillips, M.M., \& Terlevich, R. 1981, PASP, 93, 5

\bibitem [Brinchmann  et al.(2004)]{brinchmannetal2004}
          Brinchmann, J., Charlot, S., White, S.D.M., Tremonti, C., Kauffmann, G., 
          Heckman, T., \& Brinkmann, J. 2004, MNRAS, 351, 1151

\bibitem [Chiappini et al.(2005)]{chiappini2005}
          Chiappini, C., Matteucci, F., \& Ballero, S.K. 2005, A\&A, 437, 429

\bibitem [Contini et al.(2002)]{contini2002}
          Contini, T., Treyer, M.A., Sullivan, M., \& Ellis, R.S. 2002, MNRAS, 330, 75

\bibitem [Cowie et al.(1996)]{cowie1996} 
          Cowie, L.L., Songaila, A., Hu, E.M., \& Cohen, J.G. 1996, AJ, 112, 839

\bibitem [Edmunds \& Pagel(1978)]{edmundspagel1978}
          Edmunds, M.G., \& Pagel, B.E.J. 1978, MNRAS, 185, 77P

\bibitem [Esteban et al.(2004)]{estebanetal2004}
          Esteban, C., Peimbert, M., Carc\'{i}a-Rojas, J., Ruiz M.T., 
          Peimbert, A., \& Rodr\'{i}guez M. 2004, MNRAS, 355, 229

\bibitem [Izotov et al.(2006)]{izotov2006}
          Izotov, Y.I., Stasi\'{n}ska, G., Meynet, G., Guseva, N.G., \& Thuan, T.X.. 2006, A\&A, 448, 955

\bibitem [Gavil\'{a}n et al.(2005)]{gavilanetal2005}
          Gavil\'{a}n,M.,  Buell, F., \&   Moll\'{a}, M.2005, A\&A, 432, 861

\bibitem [Gavil\'{a}n et al.(2006)]{gavilanetal2006}
          Gavil\'{a}n,M.,  Moll\'{a}, M., \& Buell, F. 2006, A\&A, 450, 509

\bibitem [Henry et al.(2000)]{henryetal2000} 
          Henry, R.B.C., Edmunds, M.G., \& K\'{o}ppen, J. 2000, ApJ, 541, 660

\bibitem [Kauffmann et al.(2003)]{kauffmannetal2003} 
          Kauffmann, G., Heckman, T.M., Tremonti, C., \& et al.
          2003, MNRAS, 346, 1055

\bibitem [L\'{o}pez-S\'{a}nchez \& Esteban(2010)]{lopezsanchez2010}
          L\'{o}pez-S\'{a}nchez, \'{A}.R.,  \& Esteban, C. 2010, A\&A, 517, A85

\bibitem [Marigo(2001)]{marigo2001}
          Marigo P. 2001, A\&A, 370, 194

\bibitem [Matteucci \& Tosi(1985)]{matteucci1985}
          Matteucci, F., \& Tosi, M. 1985, MNRAS, 217, 391

\bibitem [Moustakas et al.(2010)]{moustakasetal2010}
          Moustakas, J., Kennicutt, R.C., Tremonti, C.A., Dale, D.A., 
          Smith, J.-D.T., \& Calzetti, D. 2010, ApJS, 190, 233

\bibitem [Pilyugin(1992)]{pilyugin1992}
          Pilyugin, L.S. 1992, A\&A, 260, 58

\bibitem [Pilyugin(1993)]{pilyugin1993}
          Pilyugin, L.S. 1993, A\&A, 277, 42

\bibitem [Pilyugin(1999)]{pilyugin1999}
          Pilyugin, L.S. 1999, A\&A, 346, 428

\bibitem [Pilyugin \& Thuan(2005)]{pilyuginthuan2005} 
          Pilyugin, L.S., \& Thuan , T.X. 2005, ApJ, 631, 231

\bibitem [Pilyugin et al.(2003)]{pilyuginetal2003} 
          Pilyugin, L.S., Thuan, T.X., \& V\'{\i}lchez, J.M. 2003, A\&A, 397, 487

\bibitem [Pilyugin et al.(2004)]{pilyuginetal2004} 
          Pilyugin, L.S., V\'{\i}lchez, J.M., \& Contini, T. 2004, A\&A, 425, 849

\bibitem [Pilyugin et al.(2010)]{pilyuginetal2010} 
          Pilyugin, L.S., V\'{\i}lchez, J.M., \& Thuan , T.X. 2010, ApJ, 720, 1738

\bibitem [Renzini \& Voli(1981)]{renzinivoli1981}
          Renzini, A., \& Voli, M. 1981, A\&A, 94, 175

\bibitem [Richer \& McCall(2008)]{richer2008}
          Richer, M.G., \& McCall, M.L. 2008, ApJ, 684, 1190

\bibitem [Romano et al.(2005)]{romanoetal2005}
          Romano, D., Chiappini C., Matteucci, F. \& Tosi, M. 2005, A\&A, 430, 491

\bibitem [Thuan et al.(2010)]{thuanetal2010}
          Thuan, T.X., Pilyugin, L.S., \& Zinchenko, I.A. 2010, ApJ, 712, 1029

\bibitem [Tremonti et al.(2004)]{tremontietal2004} 
          Tremonti, C.A., Heckman, T.M., Kauffmann, G., \& et al. 2004, ApJ, 613, 898

\bibitem [van den Hoek \& Groenewegen(1997)]{vandenhoek1997} 
          van den Hoek, L.B., \& Groenewegen, M.A.T. 1997, A\&AS, 123, 305

\bibitem [York et al.(2000)]{yorketal2000}
          York, D.G., et al. 2000, AJ, 120, 1579

\end{thebibliography}
\end{document}